%% file: psvmSDR.tex
\documentclass[12pt]{article}
\usepackage{jss}
\usepackage{amsmath}
\usepackage{graphicx}
\usepackage{natbib}
\usepackage{url} % not crucial - just used below for the URL 

\newcommand{\blind}{0}

\usepackage{amsmath,bm}
\let\proglang=\textsf
\let\code=\texttt

\usepackage{orcidlink,thumbpdf,lmodern}
\usepackage{framed}
\usepackage{amsthm,amssymb,amsmath,dsfont, mathtools}
\usepackage{lineno}
\usepackage{ctable}
\usepackage{multirow}
\usepackage{algorithm, algorithmicx}
\usepackage{setspace}
\usepackage{hyperref}
\usepackage{mathrsfs}
\usepackage{algpseudocode}
\usepackage{bbm}
\usepackage{graphicx}
\renewcommand{\baselinestretch}{1.5}
\usepackage{xcolor}
\usepackage{caption}
\usepackage{subcaption}
\usepackage{kotex}
\usepackage[utf8]{inputenc}
\usepackage [english]{babel}
\usepackage{enumitem}
\setlist[enumerate]{itemsep=0mm}

\usepackage{booktabs,array}
\newcolumntype{N}{@{}m{0pt}@{}}%a fix for array package

\newcommand {\vv}[1]{\mbox{vec}}

\def\argmin{\mathop{\rm argmin}}

\newcommand{\bA}{\mathbf{A}}
\newcommand{\bB}{\mathbf{B}}

\newcommand{\bM}{\mathbf{M}}

\newcommand{\bI}{\mathbf{I}}

\newcommand{\bx}{\mathbf{x}}
\newcommand{\br}{\mathbf{r}}

\newcommand{\bV}{\mathbf{V}}

\newcommand{\bX}{\mathbf{X}}

\newcommand{\bz}{\mathbf{z}}

\newcommand{\sumi}{\sum_{i=1}^{n}}

\newcommand{\cs}{\mathcal{S}_{Y\mid \bX}}

\newcommand{\bbeta}{\boldsymbol{\beta}}

\newcommand{\bmu}{\boldsymbol{\mu}}

\newcommand{\bSigma}{\boldsymbol{\Sigma}}

\begin{document}

\def\spacingset#1{\renewcommand{\baselinestretch}%
{#1}\small\normalsize} \spacingset{1}

%%%%%%%%%%%%%%%%%%%%%%%%%%%%%%%%%%%%%%%%%%%%%%%%%%%%%%%%%%%%%%%%%%%%%%%%%%%%%%

\if0\blind
{
  \title{\bf The \proglang{R} package \pkg{psvmSDR}: A Unified Algorithm for Sufficient Dimension Reduction via Principal Machines}
  \author{Jungmin Shin\\  
    Department of Statistics, Korea University\\
    Seung Jun Shin \\
    Department of Statistics, Korea University\\
    and \\
    Andreas Artemiou\\
    Department of Information Technologies, University of Limassol\\}
  \maketitle
} \fi

\if1\blind
{
  \bigskip
  \bigskip
  \bigskip
  \begin{center}
    {\LARGE\bf Title}
\end{center}
  \medskip
} \fi

\bigskip
\begin{abstract}
Sufficient dimension reduction (SDR), which seeks a lower-dimensional subspace of the predictors containing regression or classification information has been popular in a machine learning community. In this work, we present a new \proglang{R} software package \pkg{psvmSDR} that implements a new class of SDR estimators, which we call the principal machine (PM) generalized from the principal support vector machine (PSVM). The package covers both linear and nonlinear SDR and provides a function applicable to realtime update scenarios. The package implements the descent algorithm for the PMs to efficiently compute the SDR estimators in various situations. This easy-to-use package will be an attractive alternative to the \pkg{dr} \proglang{R} package that implements classical SDR methods.
\end{abstract}

\noindent%
\vfill

\newpage
\spacingset{1.75} % DON'T change the spacing!

\section[Introduction]{Introduction} \label{sec:INTRO}
Dimension reduction is essential in modern applications in statistics and machine learning as the data size grows. 
In this regard, the sufficient dimension reduction \citep[SDR,][]{Li1991::sir,Cook1998::regr} has gained great popularity in supervised learning problems.
For a given pair of univariate response $Y$ and $p$-dimensional predictor $\bX$, SDR assumes 
\begin{equation} \label{eqn:linear_model}
Y \perp \mathbf{X} \mid \mathbf{B}^{\top} \mathbf{X},     
\end{equation}
where $\perp$ denotes statistical independence. Under \eqref{eqn:linear_model}, SDR seeks the minimal space spanned by $\bB \in \mathbb{R}^{p \times d}$, called the central subspace, and denoted by $\cs$ that contains a complete regression information of $Y$ on $\bX$. We assume $\cs = \mbox{span}\{\bB\}$ in \eqref{eqn:linear_model}. The dimension of $\cs$, $d$ is also an important quantity to be estimated from the data. 

There are numerous SDR estimators to estimate $\cs$. Earliest proposals are based on an inverse moment, such as sliced inverse regression \citep[SIR,][]{Li1991::sir}, sliced average variance estimates \citep[SAVE,][]{Cook1991::save}, and principal Hessian directions \citep[pHd,][]{Li1992::phd}. The \pkg{dr} package in software \proglang{R} \citep{weisberg2015dr} concisely implements all of these methods and, to the best of our knowledge, the only \proglang{R}-package available for SDR. 

The assumption \eqref{eqn:linear_model} is often called a linear SDR. \citet{Cook2007::nonlinear} generalized \eqref{eqn:linear_model}, and proposed a nonlinear SDR that pursues a function $\phi: \mathbb{R}^p \rightarrow \mathbb{R}^d$ satisfying
\begin{equation}
\label{eqn:nonlinear_model}
Y \perp \mathbf{X} \mid \phi(\mathbf{X}).
\end{equation}
Nonlinear SDR methods have been developed by directly extending the classical inverse methods to Reproducing Kernel Hilbert Space \citep[RKHS,][]{aronszajn1950theory}. However, there is no hands-on \proglang{R}  package that implements them.

Meanwhile, \citet{Li2011::psvm} proposed the principal support vector machine (PSVM), a unified framework for both linear and nonlinear SDR, by connecting the SDR problem to the support vector machine \citep[SVM,][]{vapnik1999nature}. PSVM is known to outperform the classical SDR methods mentioned above. More importantly, PSVM can be easily generalized to other supervised machine learning approaches other than the SVM with the hinge loss. We call the class of SDR methods that generalizes the PSVM, the principal machine (PM).  

In this article, we develop an \proglang{R} package \pkg{psvmSDR} that renders a simple gradient descent algorithm to compute a wide variety of the PMs in a unified framework. 
The package offers two main functions \code{psdr()} and \code{npsdr()} that compute a wide variety of PMs for linear and nonlinear SDR, respectively. In addition, it also has a function \code{rtpsdr()} that executes the realtime SDR estimator for streamed data. 

We highlight the main advantages of \pkg{psvmSDR} package in the following.

\begin{itemize}
\item It can solve both linear and nonlinear SDR problems in a unified framework using a simple and efficient gradient descent algorithm. 
\item It can be applied to a binary classification, where most inverse-moment-based approaches suffer.
\item It can compute a PM estimator with a user-specified arbitrary convex loss, adding flexibility to the methods. 
\item  It can handle the streamed data by directly updating the SDR estimator without storing the entire data.  
\end{itemize}

The package is now available from CRAN at \url{https://cran.r-project.org/web/packages/psvmSDR} and can be installed from \proglang{R} with the following command :
\begin{Code}
 R> install.packages("psvmSDR")
\end{Code}

The rest of the article is organized as follows. In Section \ref{sec:PSDR}, we introduce the principal machine (PM) under a linear,  nonlinear, and realtime SDR framework along with the corresponding functions \code{psdr()}, \code{npsdr()}, and \code{rtpsdr()}. In Section \ref{sec:optim}, we describe computational details about the package, and its efficiencies are investigated. In Section \ref{sec:overview}, we overview the package structure and explain how to implement the functions with examples. We conclude with a summary and discussion in Section \ref{sec:CON}.

\section{Principal machines: generalization of PSVM} \label{sec:PSDR}
We start by introducing the PSVM \citep{Li2011::psvm}. For a given sequence of $c_1 < c_2 < \cdots < c_h$, the PSVM minimizes the following objective function in the population level: 
\begin{equation} \label{eqn:model}
    (\alpha_{k}, \bbeta_{k}) = \argmin_{\alpha, \bbeta} \bbeta^T \bSigma \bbeta + \lambda \E \left[\tilde{Y}_k \{\alpha + \bbeta^{\top} (\bX - \bmu)\}\right]_+, \quad k = 1, 2, \cdots, h
\end{equation}
where $\bmu = \E(\bX)$, $\bSigma = \COV(\bX)$, and $[u]_+ = \max\{0, u\}$. Here $\tilde{Y}_{k}$ is a pseudo-binary response, taking 1 if $Y < c_k$ and $-1$ otherwise, and a positive constant $\lambda$ is a cost parameter whose choice is not overly sensitive for estimating $\cs$. \citet{Li2011::psvm} showed the unbiasedness of $\bbeta_{k}$, i.e., $\bbeta_{k} \in \cs, \forall k = 1, 2, \cdots, h$.

It is important to note that convexity of the objective function in \eqref{eqn:model} is the only requirement for the PSVM to be unbiased, and this naturally leads to a generalized version of the PSVM, which we call the principal machine \citep[PM,][]{shin2024concise}. PMs are categorized into two types: one is response-based PM (RPM), and the other is loss-based PM (LPM). The RPM obtains multiple solutions by perturbing the pseudo-response $\tilde Y_k$, while the loss function remains unchanged. The PSVM belongs to this category. 
On the other hand, the LPM calculates multiple solutions by changing the loss functions, while the pseudo-response $Y_k$ is fixed as $Y$ for all $k$. 
The principal weighted support vector machine \citep[PWSVM,][]{shin2017principal} and the principal quantile regression \citep[PQR,][]{wang2018principal} are the earliest proposals of the LPM.  

\subsection{Linear principal machines} \label{sec:LPSDR}
The linear PM solves
\begin{align} \label{eqn:pm}
(\alpha_{k}, \bbeta_k) = \argmin_{\alpha, \bbeta} \bbeta^\top \bSigma\bbeta + \lambda \E \left[L_{k} (\tilde Y_k, f(\bX)\} \right], \qquad k = 1, 2, \cdots h,
\end{align}
where $f(\bX) = \alpha + \bbeta^\top (\bX - \bmu)$ and $L_k(y, f)$ denotes a convex loss function of a margin, $yf$ for a binary $y \in \{-1, 1\}$ or residual $y - f$ for a continuous $y \in \mathbb{R}$. The RPM solves \eqref{eqn:pm} for different values of $\tilde Y_k$ over $k$ to obtain multiple solutions while keeping $L_k$ fixed as say, $L$. On the other hand, the LPM  solves it with different loss functions $L_k$ while a pseudo-response $\tilde Y_k$ remains fixed as the original response $Y$. For example, PWSVM, the first proposal of the LPM solves \eqref{eqn:pm} for $L_k(y, f) = \pi_k(y)[1-yf]_+$ where $\pi_k(y) = c_k$ if $y = 1$ and $1 - c_k$ otherwise for a given $c_k \in (0,1)$ while $\tilde Y$ is fixed as $Y \in \{-1, 1\}$. The PWSVM is proposed for SDR in a binary classification, where the PSVM fails when $d>1$. 

Given $(y_i, \bx_i), i = 1, 2, \cdots, n$, the sample counter part of \eqref{eqn:pm} is
\begin{equation}
\label{eqn:sample_model}
(\hat \alpha_k, \hat \bbeta_k) = \argmin_{\alpha, \bbeta} \bbeta^{\top} \widehat \bSigma \bbeta + \frac{\lambda}{n} \sumi  L_k\left(\tilde y_{k,i}, \alpha + \bbeta^\top \bz_i \right), \qquad k = 1, 2, \cdots h,
\end{equation}
where $\bz_i = \bx_i - \sumi \bx_i/n$ and $\widehat \bSigma = n^{-1}\sumi \bz_i \bz_i^\top$ denotes a sample covariance matrix. The PM working matrix is then 
\begin{equation} \label{eqn:working_matrix}
\widehat{\mathbf{M}} = \sum_{k=1}^h \hat{\boldsymbol{\beta}}_{k} \hat{\boldsymbol{\beta}}_{k}^{\top},
\end{equation}
and its first $d$ eigenvectors of $\widehat \bM$ estimate $\bB$ under \eqref{eqn:linear_model}, where $d$ denotes the dimension of $\cs$, often called as structure dimension.
The \pkg{psvmSDR} package offers \code{psdr()} function to compute the linear PM estimates with a given loss function. 

In the linear SDR, it is also crucial to estimate the structure dimension  $d$. The linear PM employs the following BIC-type criterion proposed by \citet{Li2011::psvm} to estimate $d$:
\begin{equation}\label{eqn:structure dimension}
\hat{d}=\underset{d \in\{1, \cdots, p\}}{\operatorname{argmax}} \sum_{j=1}^{d} v_{j}-\rho \frac{d \log n}{\sqrt{n}} v_{1},
\end{equation}
where $v_{1} \geq \cdots \geq v_{p} (p > d)$ are eigenvalues of $\widehat{\mathbf{M}}$ in \eqref{eqn:working_matrix}, with $\rho$ being a hyperparameter that can be chosen in a data-adaptive manner. The \pkg{psvmSDR} package contains \code{psdr\_bic()} function for estimating $d$ based on \eqref{eqn:structure dimension}.

There are numerous PMs by employing a variety of convex loss function in supervised learning problems. 
For RPMs, the principal logistic regression with negative log-likelihood loss \citep[PLR,][]{shin2017penalized}, the principal $L_q$-SVM with $L_q$ hinge loss \citep[$L_q$-PSVM,][]{artemiou2016sufficient}, and the principal least square SVM with the squared loss \citep[PLSSVM,][]{artemiou2021real} are proposed.
For LPMs, \citet{wang2018principal} suggested the principal quantile regression (PQR) with the check loss, \citet{kim2019principal} proposed the principal weighted logistic regression (PWLR), \citet{soale2022sufficient} introduced the principal asymmetric least square regression (PALS) with the asymmetric least square loss, and \citet{jang2023principal} proposed the principal weighted least square SVM (PWLSSVM).

Table \ref{tbl:loss} provides a complete list of PMs implemented in \pkg{psvmSDR} with the corresponding loss functions. 

\input{tbl_loss.tex}

\subsection{Kernel principal machines for nonlinear SDR} \label{sec:NLPSDR}

A nonlinear generalization of \eqref{eqn:pm} under \eqref{eqn:nonlinear_model} is
\begin{equation}\label{model:non_pm_population}
(\alpha_{0,k}, g_{0,k}) = \argmin_{\alpha \in \mathbb{R}, g \in \mathcal{H}} \VAR \{\psi(\bX)\}+\lambda \E\left[ {L}_k \left\{ \widetilde{Y}_{k}, f(\bX) \right\} \right], \qquad k = 1, 2, \cdots, h,
\end{equation}
where $f(\bX)=\alpha+g(\bX)-\E\{g(\bX)\}$ with $g$ being a function in a Hilbert space $\mathcal{H}$ of functions of $\bX$. 
Let $(\alpha_{0,k}, g_{0,k})$ be the minimizer of \eqref{model:non_pm_population}, then $g_{0,k}(\bX)$ is necessarily a function of the sufficient predictor $\phi(\bX)$ in \eqref{eqn:nonlinear_model} and its unbiasedness for the nonlinear SDR is established by \citet{Li2011::psvm}.

Employing the reproducing kernel Hilbert space (RKHS) generated by a positive definite kernel $K(\cdot, \cdot)$ for the space of $g$, we have the following finite-dimenional basis representation of $g$:
\begin{align}\label{eqn:fx}
g(\bx) = \sum_{j=1}^b \beta_j \left\{\psi_{j}(\mathbf{x})- \bar \psi_j \right\}
\end{align}
with 
\begin{align}\label{eqn:psi_gen}
\psi_j(\mathbf{x})=\left\{\mathbf{k}(\mathbf{x})\right\}^{\top} \mathbf{q}_j / \lambda_j, \quad j=1, \cdots b, \qquad \mbox{and} \qquad \bar \psi_j = \sum_{i=1}^{n}\psi_{j}\left(\mathbf{x}_{i}\right)/n,
\end{align}
where $\mathbf{k}(\cdot)=\{K(\cdot, \bx_i) : i=1,\ldots,n\}^{\top}$, and $\mathbf{q}_j$ and $\lambda_j$ are the $j$th leading eigenvector and eigenvalue of the centered kernel matrix $\mathbf{K} = \{K_{ij}\} \in \mathbb{R}^{n\times n}$ with $K_{ij} = K(\bx_i, \bx_j) - \sum_{k = 1}^n K(\bx_k, \bx_j)/n$. We refer Section 6 of \citet{Li2011::psvm} for the theoretical justification for \eqref{eqn:fx} and \eqref{eqn:psi_gen}.  In \pkg{psvmSDR}, we employ the radial/Gaussian kernel and and set $b = n/3$ as recommended by \citet{Li2011::psvm}. 
 
By employing \eqref{eqn:fx}, the sample counter part of \eqref{model:non_pm_population} is 
\begin{equation}\label{eqn:sample_nonlinear_pm}
(\hat \alpha_k, \hat \bbeta_k) = \argmin_{\alpha, \bbeta} {\bbeta}^{\top} \hat \bSigma {\bbeta}+\frac{\lambda}{n} 
   \sum_{i=1}^{n} L_{k} \left( \widetilde{y}_{k,i}, \alpha + {\bbeta}^{\top} \bz_{i} \right), \qquad k = 1, 2, \cdots, h,   
\end{equation}
where $\bz_i = \sum_{j=1}^b \psi_{j}(\mathbf{x}_i)- \bar \psi_j$ and $\hat \bSigma = n^{-1} \sumi \boldsymbol{\bz}_i \boldsymbol{\bz}_i^{\top}$. We note that this kernel PM in \eqref{eqn:sample_nonlinear_pm} is also linear in the parameter, exactly the same as the linear PM \eqref{eqn:sample_model}. 
For a given $\bx$, one can compute the sufficient predictors, $\hat{\phi}(\bx)=\widehat{\bV}^{\top}\left\{ \psi_1(\bx), \ldots,  \psi_b(\bx)\right\}^{\top}$, where $\widehat{\bV}$ denotes the $d$-leading eigenvectors of $\sum_{k=1}^h \hat{\bbeta}_{k} \hat{\bbeta}_{k}^{\top}$. The \pkg{psvmSDR} package offers \code{npsdr()} function to compute the nonlinear PM estimates.

\subsection{Principal least square machines for realtime SDR} \label{sec:rtPSDR}
When data is collected in a streamed fashion, it is prohibitive to use all data to compute the SDR estimator due to memory constraints. Therefore, it is important to develop real-time SDR algorithms that directly update the SDR estimators without storing entire data. In this regard, \citet{artemiou2021real} proposed PLSSVM which solves \eqref{eqn:pm} with the squared loss $L(\tilde y, f) = (1 - \tilde y f)^2$.

Let's consider the following scenario. In addition to the currently available (old) data $\mathbb{D}_{\texttt{O}} = (y_i, \bx_i), i = 1, 2, \cdots, n$, suppose $m$ additional (new) data $\mathbb{D}_{\texttt{N}} = (y_i, \bx_i), i = n+1, \cdots, n+m$ arrives. Let $\mathbb{D}_{\texttt{W}} = \mathbb{D}_{\texttt{O}} \cup \mathbb{D}_{\texttt{N}}$ denote the whole data. The subscripts $\texttt{O}, \texttt{N}$, and $\texttt{W}$ are used to denote the quantities related to old, new, and whole data, respectively.  
\citet{artemiou2021real} showed that the PLSSVM solution for the whole data, ${\br}_{\texttt{W}} = (\hat{\alpha}_{\texttt{W}},\hat {\bbeta}_{\texttt{W}}^\top)^\top\in \mathbb{R}^{p+1}$, is given by
\begin{align}\label{eqn:rtpsdr_coef}
\br_{\texttt{W}} = \{\bI - \bA_{\texttt{O}}^{-1} \bB_{\texttt{N}} (\bI + \bA_{\texttt{O}}^{-1} \bB_{\texttt{N}})^{-1} \} ( \br_{\texttt{O}} + \bA_{\texttt{O}}^{-1}\mathbf{c}_{\texttt{N}} ),
\end{align}
where $\br_{\texttt{O}}$ is the solution for $\mathbb{D}_{\texttt{O}}$, and $\bA_{\texttt{O}}^{-1} \in \mathbb{R}^{{(p+1)}\times {(p+1)}}$ is also computable from $\mathbb{D}_{\texttt{O}}$, while $\bB_{\texttt{N}} \in \mathbb{R}^{{(p+1)}\times {(p+1)}}$ and $\mathbf{c}_{\texttt{N}} \in \mathbb{R}^{p+1}$ are computable from $\mathbb{D}_{\texttt{N}}$. That is, one can directly update $\br_{\texttt{W}}$ from $\br_{\texttt{O}}$ without storing entire $\mathbb{D}_{\texttt{O}}$, but $\bA_{\texttt{O}}$ only. We remark that $\bA_{\texttt{W}}$ can be directly updated from $\bA_{\texttt{O}}$ and the new data, and its computational complexity does not depend on the size of $\mathbb{D}_\texttt{O}$. We refer Section 3 of \citet{artemiou2021real} for the exact forms of these quantities.

The \pkg{psvmSVM} package offers \code{rtpsdr()} that implements the realtime SDR using PLSSVM for a continuous response as well as PWLSSVM by \citet{jang2023principal} for a binary response.

\section{Computation}\label{sec:optim}
In this section, we describe computational details about \code{psdr()} and \code{npsdr()} in \pkg{psvmSDR}.
Slightly abusing notation as $\bbeta^\top = (\alpha, \bbeta^\top)$ and $\bz_i^\top = (1, \bz_{i}^\top)$, the objective functions \eqref{eqn:sample_model} and \eqref{eqn:sample_nonlinear_pm} are then rewritten as
\begin{align} \label{eqn:obj}
L(\bbeta) = \bbeta^\top \mbox{Diag}\{0, \widehat \bSigma\} \bbeta + \frac{\lambda}{n}\sumi L_k\left(\tilde y_{k,i}, \bbeta^\top \bz_i \right).
\end{align}
We propose the coordinatewise gradient descent (CGD) algorithm to minimize $L(\bbeta)$ in \eqref{eqn:obj}, which updates $\bbeta$ coordinatewisely as follows until converge:
\begin{align*}
\beta_j & \leftarrow \beta_j - \eta \frac{\partial L(\bbeta)}{\partial{\beta_j}}, ~~ j= 0, 1, 2, \cdots, p,
\end{align*}
where $\eta > 0$ denotes a learning rate determined by the user. 
We remark that some loss functions, such as the hinge loss and check loss are not theoretically differentiable at most finite number of points while numerically differentiable. The two main functions \code{psdr()} and \code{npsdr()} in \pkg{psvmSDR} solve linear and kernel PMs via this CGD algorithm, respectively, and cover a wide variety of PM estimators by simply changing the loss functions (via \code{`loss'} argument) as listed in Table \ref{tbl:loss}. The CGD algorithm is easily extended to any user-defined convex loss function, since the corresponding derivative can be readily evaluated numerically. Both functions take the name of the user-defined-loss function object as the \code{loss} argument which brings further flexibility to the PM estimators. 

\begin{figure}[!ht]
     \begin{subfigure}[b]{0.45\textwidth}
         \centering
        \includegraphics[width=0.95\textwidth]{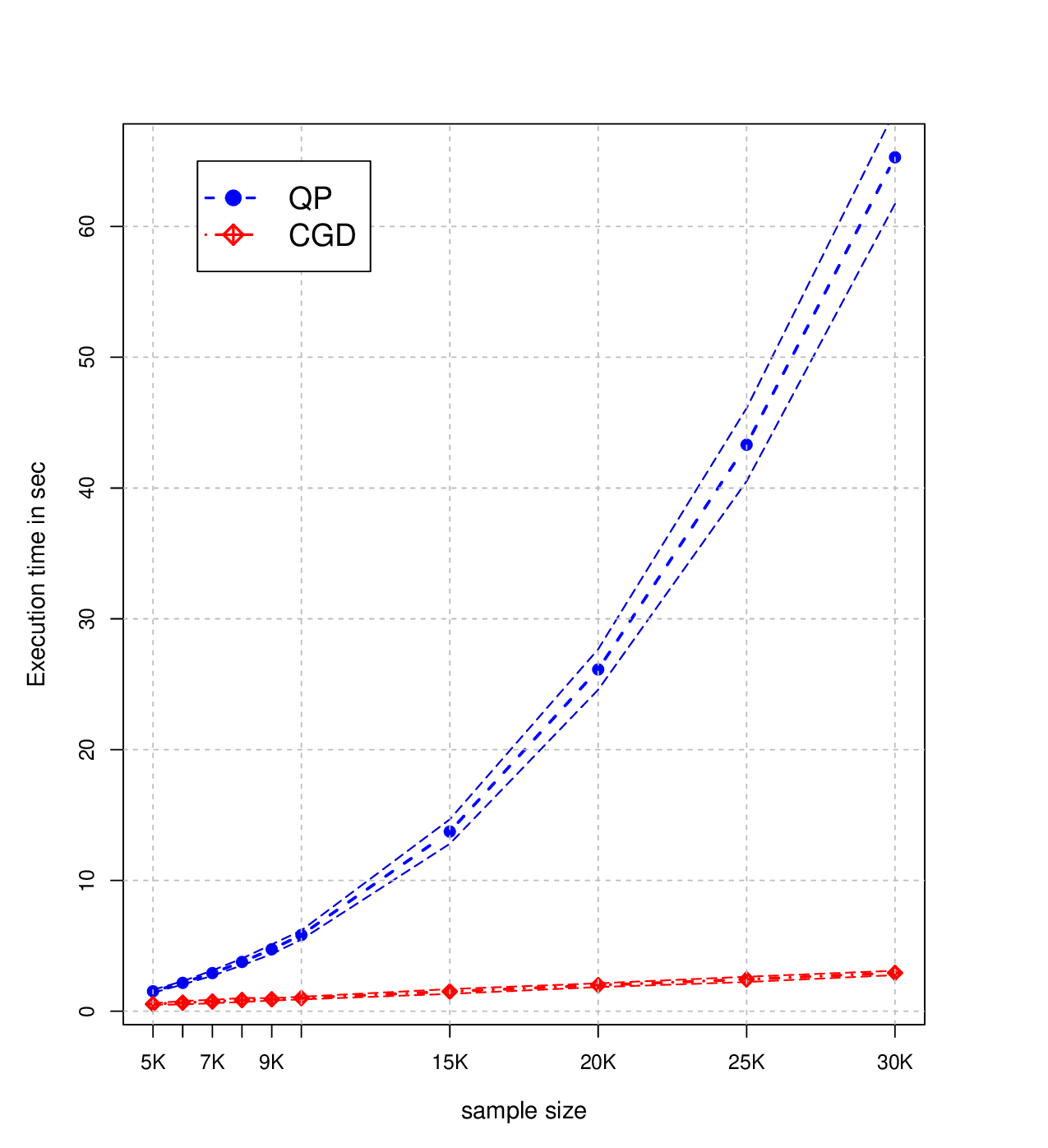}
        \caption{Linear PSVM}
        %\label{fig:linear_time}
     \end{subfigure}
     \hfill
     \begin{subfigure}[b]{0.48\textwidth}
         \centering
        \includegraphics[width=0.95\textwidth]{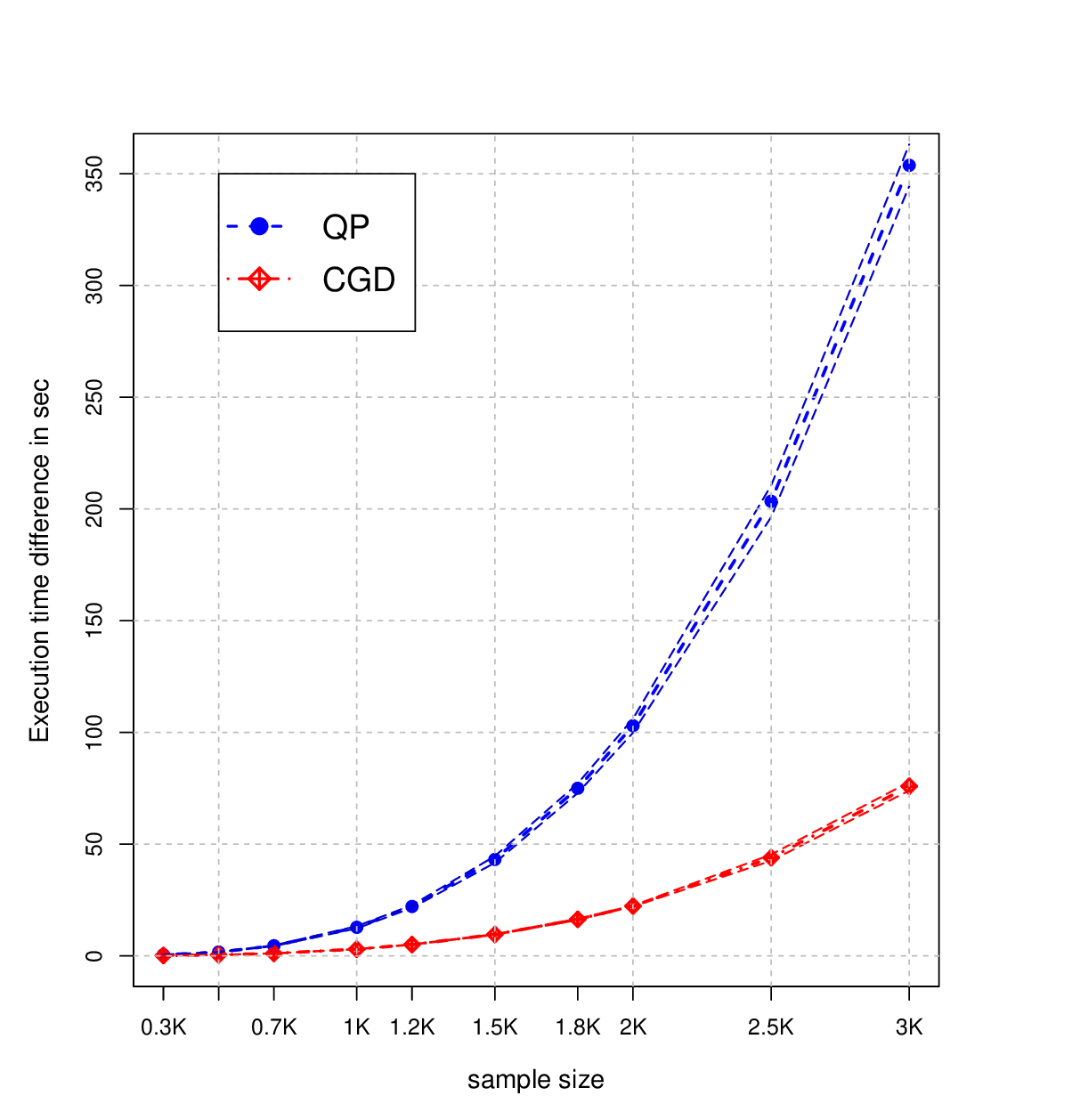}
        \caption{Kernel PSVM}
        %\label{fig:nonlinear_time}
     \end{subfigure}
        \caption{Averaged computing times for QP and CGD algorithms over 100 repetitions are depicted. The blue dashed line with asterisks is for QP and the red dashed line with diamond is for CGD algorithm. The computing time gap increases substantially for the linear and nonlinear cases as the sample size gets larger. The dashed lines present the confidence intervals for both algorithms.}
         \label{fig:comparison_methods}
\end{figure}

The time complexity of the gradient descent method is known as $\mathcal{O}(np)$ \citep{luo1992convergence, tibshirani2007cd}, where $p$ is the data dimension. Whereas, quadratic programming (QP), which is widely used in the context of solving an SVM-like problem via \proglang{R} package \pkg{kernlab}, is known as to have a polynomial rate, $\mathcal{O}(n^3p^3)$, at worst \citep{platt1998sequential, cristianini2000introduction}. To illustrate the computational efficiency of the proposed CGD algorithm, we compare its computing times for both linear and kernel PSVM estimators to \code{ipop()} function in \pkg{kernlab}, a standard QP solver for the SVM over 100 independent repetitions. A toy data is generated from the following regression model:
\begin{align}\label{model_regression_1}
y_i = x_{i1} / \left\{0.5+\left(x_{i2} + 1\right)^2\right\} + 0.2\epsilon_i,
\end{align}
where both $x_{ij}$ and $\epsilon_i$ are i.i.d. random samples from $N(0,1)$ for $i = 1, \cdots, n$ and $j = 1, 2, \cdots, 5$. Different sample sizes $n$ are considered from $5,000$ to $30,000$ for linear PSVM while from $300$ to $3,000$ for the kernel PSVM where the parameter dimension depends on $n$. Figure \ref{fig:comparison_methods} contrasts the computing times between the CGD and the QP algorithms. 
It clearly demonstrates the computational efficiency of CGD algorithm implemented in \pkg{psvmSDR}.

%\begin{figure}[!htb]
%    \centering
%    \subfigure[]{\includegraphics[width=0.40\textwidth]{comp_time_linear_vff.eps}}
%    \subfigure[]{\includegraphics[width=0.42\textwidth]{comp_time_nonlinear_new.eps}}
%        \caption{The left panel \textbf{(a)} is for \textbf{Linear PSVM} and the right \textbf{(b)} is for \textbf{Kernel PSVM}. Averaged computing times for QP and CGD algorithms over 100 repetitions are depicted. The blue dashed line with asterisks is for QP and the red dashed line with diamond is for CGD algorithm. The computing time gap increases substantially for the linear and nonlinear cases as the sample size gets larger. The dashed lines present the confidence intervals for both algorithms.}
%         \label{fig:comparison_methods}
%\end{figure}

\section[Structure]{Package overview and implementation}\label{sec:overview}
\subsection[overview]{Overview of the package \pkg{psvmSDR}}
The \pkg{psvmSDR} package is organized to allow users to run different types of PM methods in a unified fashion by calling just a few interface functions. 
Several auxiliary codes are then internally called to perform computations according to the options specified in those interface functions.
The overall design of \pkg{psvmSDR} adopts the functional object-oriented programming approach
\citep{chambers2014::R} with \code{S3} classes and methods. Every function in the package is either
a wrapper that creates a single instance of an object or a method that can be applied to a
class object.

The package can be loaded in an \proglang{R} session via:
\begin{CodeInput}
R> install.packages("psvmSDR")
R> library("psvmSDR")
\end{CodeInput}

The main functions in the package are \code{psdr()} and \code{npsdr()}. 
These are designed to be analogous to linear and nonlinear principal sufficient dimension reduction methods, respectively. 
Each function returns an \proglang{S3} object, \code{`psdr'} and \code{`npsdr'}, respectively. 
Also, compatible \proglang{S3} methods \code{plot()} and \code{print()} let the users enjoy the result of dimension reduction through familiar generic functions. 
Moreover, some explicit methods of which are related to the main functions are also provided, such as \code{psdr\_bic()} and \code{npsdr\_x()}.

A function \code{rtpsdr()} implements the principal least square SVM in a realtime fashion \citep{artemiou2021real,jang2023principal} discussed in Section~\ref{sec:rtPSDR}.

Table \ref{tbl:ft_list} summarizes the main functions along with their compatible methods.
In the following subsections,  we demonstrate the general usage of the main functions and methods of \pkg{psvmSDR} by the \code{S3} class of the function.

\input{tbl_overview.tex}

\subsection[Functions for S3 class `psdr']{Functions for \code{S3} class `\code{psdr}'}\label{sec:class_psdr}

A function \code{psdr}() provides a unified interface for all linear PMs solved by the coordinatewise gradient decent algorithm.
The following arguments are used (some are mandatory) when calling \code{psdr}() function and they must be provided in the order that follows:
\begin{Code}
 R> psdr(x, y, loss = "svm", h = 10, lambda = 1,
         eps = 1.0e-5, max.iter = 100, eta = 0.1, mtype="m", plot = FALSE)
\end{Code}
\begin{itemize}[itemsep=.1mm]
\item \code{x}: input matrix; each row is an observation vector. \code{x} should have 2 or more columns.
\item \code{y}: response vector, either can be continuous or $\{1, -1\}$-coded binary variable.
\item \code{loss}: name of loss function objects listed in Table \ref{tbl:loss}, or the name of a user-defined (convex) loss function object. 
\item \code{h}: number of slices.
\item \code{lambda}: cost parameter.
\item \code{eps}: stopping criterion of the CGD algorithm.
\item \code{max.iter}: maximum iteration number of the CGD algorithm.
\item \code{eta}: learning rate for the CGD algorithm.
\item \code{mtype}: a margin type, which is either margin ("m") or residual ("r") (See, Table~\ref{tbl:loss}). Only needed when user-defined loss is used. The default is "m".
\item \code{plot}: boolean. If \code{TRUE}, scatter plots of $Y$ and sufficient predictors are depicted. 
\end{itemize}

The main function \code{psdr()} returns an object with \code{S3} class \code{`psdr'} including
\begin{itemize}
	\item \code{evalues}: eigenvalues of the estimated working matrix \eqref{eqn:working_matrix}.
	\item \code{evectors}: eigenvectors of the estimated working matrix \eqref{eqn:working_matrix}.
\end{itemize}

To further illustrate how to use function \code{psdr()}, we reuse the toy example which is formulated in \eqref{model_regression_1} with $n=200$.
\begin{CodeChunk}
\begin{CodeInput} 
 R> set.seed(100)
 R> n <- 200; p <- 5
 R> x <- matrix(rnorm(n*p, 0, 1), n, p)
 R> y <- x[,1]/(0.5 + (x[,2] + 1)^2) + 0.2 * rnorm(n)   
\end{CodeInput} 
\end{CodeChunk}

We then can apply the PSVM \citep{Li2011::psvm} using function \code{psdr()} with calling the argument \code{loss = "svm"} (default value) as below: 
\begin{CodeChunk}
\begin{CodeInput} 
 R> obj <- psdr(x, y)
 R> print(obj)
\end{CodeInput}
\begin{CodeOutput}
$evalues
[1] 0.7802035134 0.0450979178 0.0069745442 0.0013436194 0.0002315534
$evectors
             [,1]          [,2]         [,3]        [,4]        [,5]
[1,]  0.996301948  0.0002739995 -0.006936561  0.07908355  0.03286382
[2,]  0.001962004 -0.9577841701  0.125064041 -0.10815774  0.23517339
[3,] -0.018812750 -0.2384741801  0.125542475  0.57099528 -0.77522875
[4,] -0.045008311  0.1507599374  0.767379695  0.46014120  0.41790450
[5,]  0.070702982  0.0552350897  0.616180461 -0.66654494 -0.40986542
\end{CodeOutput}
\end{CodeChunk}
The output \code{"evectors"} correctly estimate the true basis vectors $(1,0,0,0,0)^{\top}$ and $(0,1,0,0,0)^{\top}$. 
The method \code{plot.psdr()} creates the scatter plots between response and the $j$-th sufficient predictors. 
\begin{Code}
 R> plot(obj)
\end{Code}
\begin{figure}[!h]
    \centering
    \includegraphics[width=.50\linewidth]{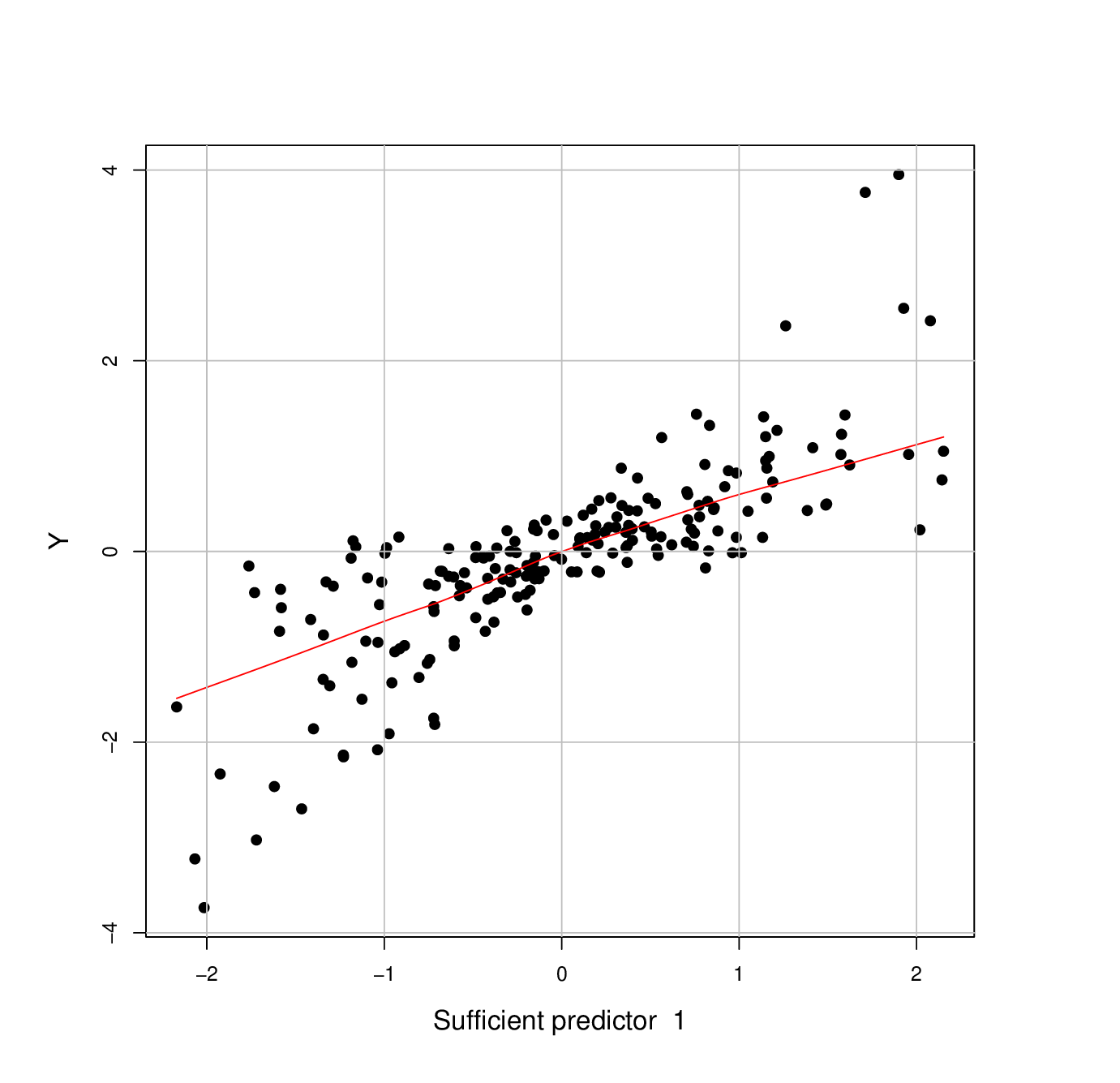}
    \caption{Scatter plots of $Y$ versus the first sufficient predictor $\hat{B}_{1}^{\top} \mathbf{X}$ from the generic method \code{plot.psdr()}.}
    \label{fig:ex1_psdr}
\end{figure}

The argument \code{obj} is the object from the function \code{psdr()} which contains information on estimated $\cs$. The number of sufficient predictors to be plotted is specified by the argument \code{d} whose default value is 1. By default, a locally weighted scatterplot smoothing (LOWESS) curve is plotted, unless \code{lowess=FALSE} is specified. The argument \code{"$\ldots$"} allows for additional arguments to be passed to generic \code{plot()} function such as \code{main}, \code{cex}, \code{col}, \code{lty}, and etc.

%\tbf{2. Linear PWSVM}

Now, we generate binary response $\tilde y_i = \operatorname{sign}(y_i)$ from \eqref{model_regression_1} to illustrate the PWSVM \citep{shin2017principal}, an example of the LPM. One can apply PWSVM by letting \code{loss = "wsvm"}. 
\begin{CodeChunk}
\begin{CodeInput} 
 R> y.binary <- sign(y)
 R> obj_wsvm <- psdr(x, y.binary, loss = "wsvm")
 R> print(obj_wsvm)
 R> plot(obj_wsvm)
\end{CodeInput}
\begin{CodeOutput}
$evalues
[1] 3.476293e-01 7.191191e-12 1.127804e-14 1.307353e-15 1.413672e-16
$evectors
            [,1]        [,2]        [,3]        [,4]          [,5]
[1,]  0.99381714  0.01602173  0.05672899  0.06954918  0.0633682181
[2,]  0.05882480  0.63307786 -0.56478786 -0.52608610  0.0003881853
[3,] -0.01789413 -0.27655680 -0.69467381  0.41136035  0.5209674569
[4,]  0.02034888 -0.48647584  0.12690122 -0.71901964  0.4794100790
[5,]  0.09018236 -0.53461511 -0.42322673 -0.17941503 -0.7033798428
\end{CodeOutput}
\end{CodeChunk}

\begin{figure}[!ht]
    \centering
    \includegraphics[width=.55\linewidth]{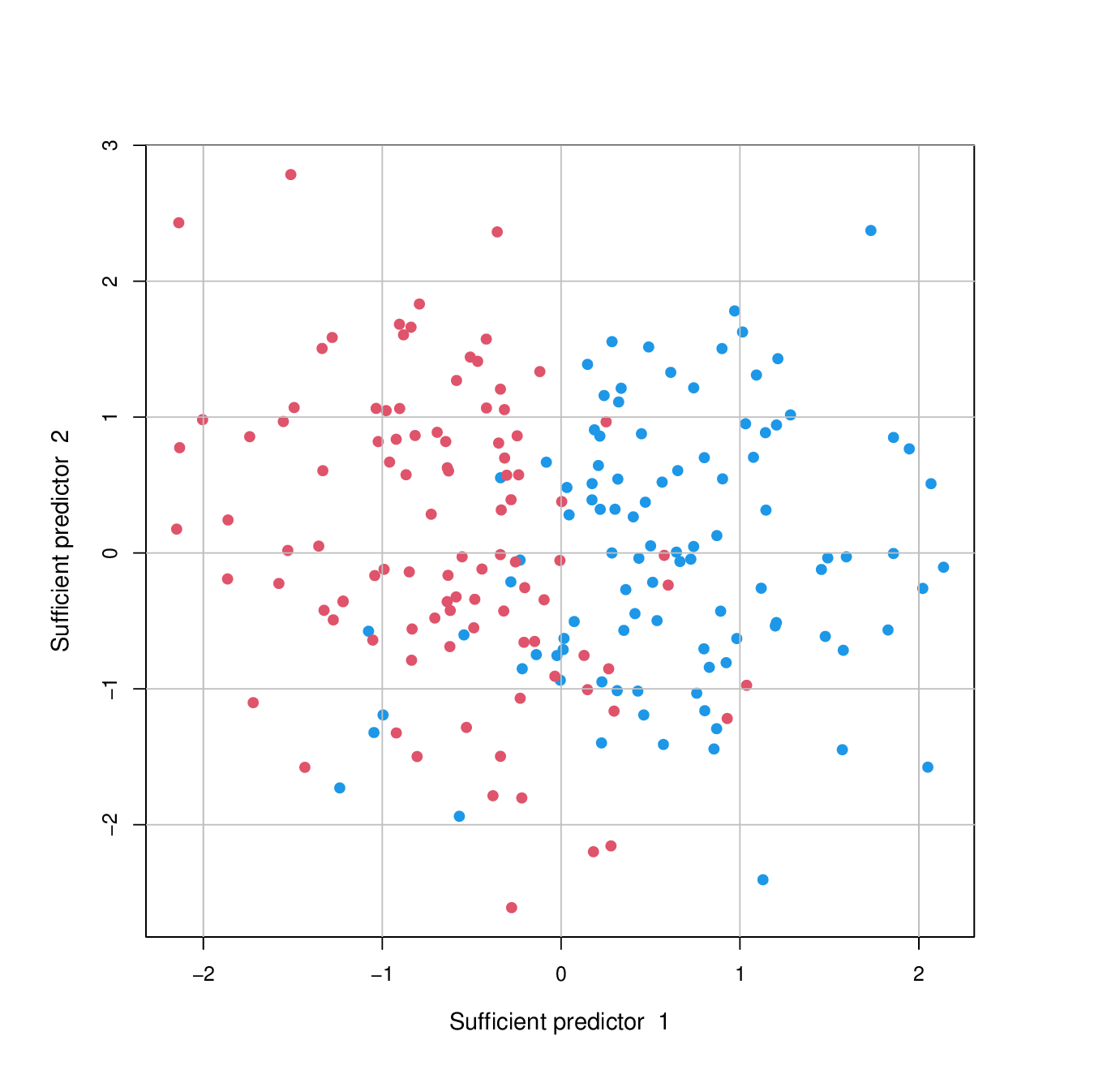}  
    \caption{A scatter plot of the predicted projections on the first and second sufficient predictors, i.e., $\hat{b}_1^{T}\bX$, $\hat{b}_2^{T}\bX$, respectively, estimated by PWSVM with \code{loss=``wsvm''}. }
    \label{fig:ex1_wsvm}
\end{figure}

Figure~\ref{fig:ex1_wsvm} visualizes the classification performance with only two sufficient predictors estimated by PWSVM.
With only two sufficient predictors, the two classes are well separated.

One of many advantages of \code{psdr()} is that it can adopt any convex user-defined loss function in addition to existing choices listed in Table \ref{tbl:loss}.
A basic syntax of the arbitrary loss function follows:
\begin{CodeChunk}
\begin{CodeInput} 
loss_name <- function(u, ...){ body of a function }
\end{CodeInput} 
\end{CodeChunk}
Argument \code{u} is a variable of a function  (any character is possible) and any additional parameters of the loss function can be specified via \code{...} argument.
For example, one can define `\code{mylogistic}' and apply it to the function \code{psdr()} by specifying \code{loss=``mylogistic"}.
\begin{CodeChunk}
\begin{CodeInput} 
 R> mylogistic <- function(u) log(1+exp(-u))
 R> obj_mylogistic <- psdr(x, y, loss="mylogistic")
 R> print(obj_mylogistic)
\end{CodeInput}
\begin{CodeOutput}
$evalues
[1] 1.556551e-01 8.926810e-03 1.407538e-03 2.607252e-04 4.640626e-05

$evectors
             [,1]          [,2]         [,3]        [,4]        [,5]
[1,]  0.996336971  0.0002783802 -0.008433507  0.07818789 -0.03358709
[2,]  0.002321387 -0.9579820189  0.125656435 -0.11231156 -0.23208110
[3,] -0.018941160 -0.2384204646  0.119983562  0.58427863  0.76617188
[4,] -0.043428997  0.1495969167  0.768896248  0.45467693 -0.42166590
[5,]  0.071150592  0.0551979934  0.615257540 -0.65814903  0.42448651
\end{CodeOutput}
\end{CodeChunk}

As aforementioned, \pkg{psvmSDR} provides a function \code{psdr\_bic()} that implements the BIC-type criterion in \eqref{eqn:structure dimension}.
Following arguments are mandatory when calling the \code{psdr\_bic()}:

\begin{Code}
R> d.hat <- psdr_bic(obj, rho = 0.01, plot = TRUE, ...)
\end{Code}

\begin{itemize}
\item \code{obj}: the output object from the main function \code{psdr()} 
\item \code{rho}: A hyperparameter in the BIC criterion.
\item \code{plot}: Boolean. If \code{TRUE}, the plot of BIC values is depicted.  
\item \code{$\cdots$}: additional arguments to be passed to generic \code{plot()}.
\end{itemize}

\begin{CodeChunk}
\begin{CodeInput}
 R> d.hat <- psdr_bic(obj, rho=0.03)
 R> print(d.hat)
\end{CodeInput}
\begin{CodeOutput}
[1] 0.7714345 0.8077633 0.8059689 0.7985434 0.7900059
\end{CodeOutput}
\end{CodeChunk}

\begin{figure}[!ht]
     \centering
        \includegraphics[width = 0.55\linewidth]{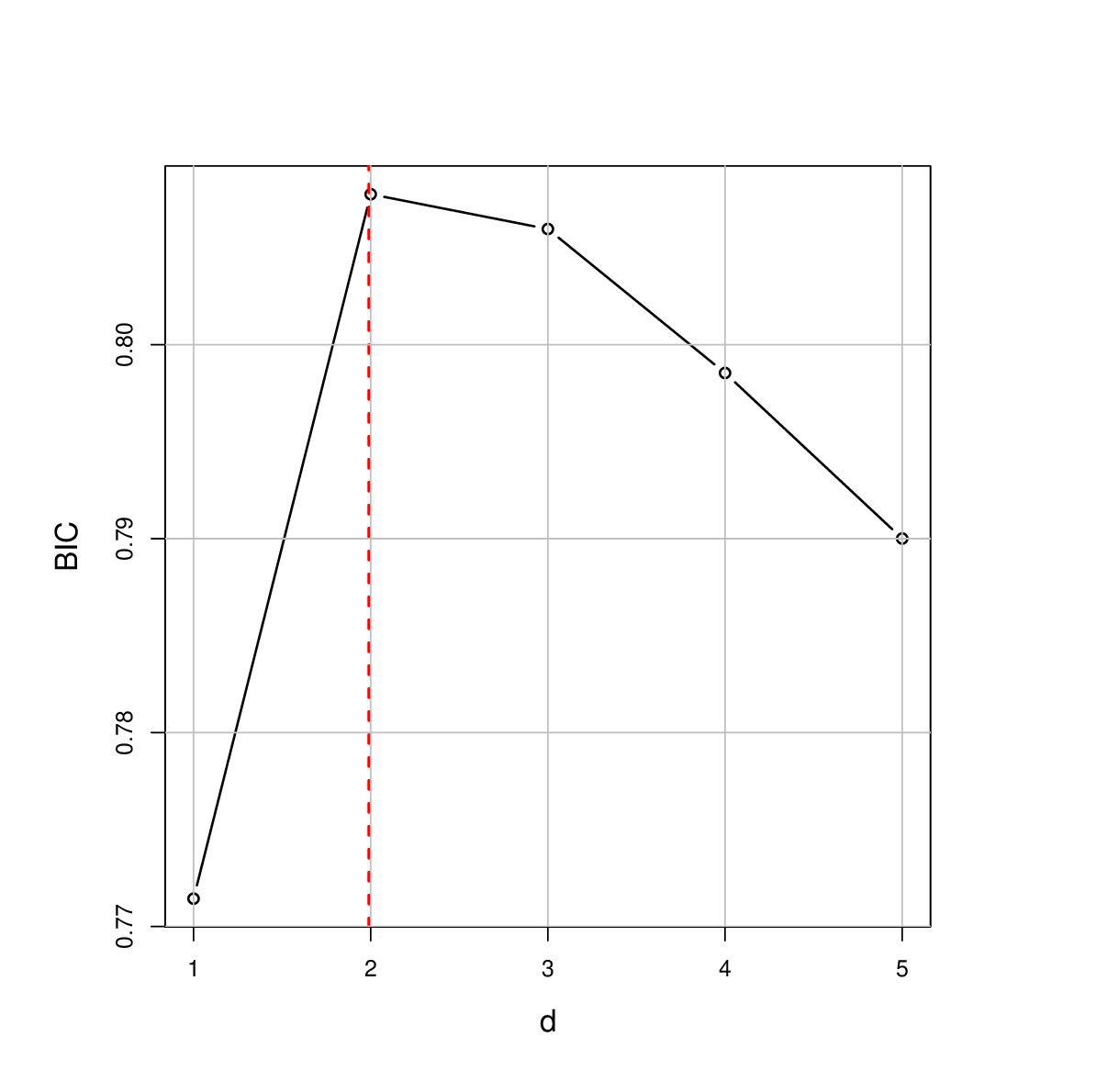}
        \caption{BIC values computed from \code{psdr\_bic()} with \code{plot = TRUE}.}
        \label{fig:plot and Gn}
\end{figure}

\subsection[Functions for class S3 `npsdr']{Functions for \code{S3} class \code{`npsdr'}}\label{sec:class_npsdr}
The function \code{npsdr()} is a nonlinear version of \code{psdr} that implements kernel PMs. 
A function \code{npsdr()} receives the following arguments:
\begin{Code}
 R> npsdr(x, y, loss="svm", h = 10, lambda = 1, b = floor(n/3), 
          eps = 1.0e-5, mtype="m", max.iter = 100)
\end{Code}
All input arguments are identical to those for \code{psdr()} except \code{b} which denotes the number of basis functions on RKHS in \eqref{eqn:psi_gen}. The function \code{npsdr()} returns an object with \code{S3} class \code{"npsdr"} that contains the following.
\begin{itemize}%[itemsep=.2mm]
	\item \code{evalues}: eigenvalues of the estimated working matrix.
	\item \code{evectors}: eigenvectors of the estimated working matrix.
        %\item {\code{obj.psi}}: 
\end{itemize}
The kernel information is stored internally and is not explicitly returned unless forced to print the object \code{obj.psi}.

To further illustrate how to use function \code{npsdr()}, we use the following model used in \cite{Li2011::psvm}:
\begin{align}\label{model_regression_2}
y_i =  0.5\left(x_{i1}^2 + x_{i2}^2 \right)^{1/2}   \mathrm{log}\left(x_{i1}^2 + x_{i2}^2 \right) + 0.2\epsilon_i, \quad i=1,\ldots,200.
\end{align}

Following code is for demonstrating kernel PSVM using function \code{npsdr()}.
\begin{CodeChunk}
\begin{CodeInput}
 R> set.seed(100)
 R> n <- 200; p <- 5
 R> x <- matrix(rnorm(n*p, 0, 1), n, p)
 R> y <- 0.5*sqrt((x[,1]^2+x[,2]^2))*(log(x[,1]^2+x[,2]^2))+ 0.2*rnorm(n)
 R> obj_kernel <- npsdr(x, y, max.iter = 200, eta = 0.8, plot = FALSE)
 R> print(obj_kernel)
\end{CodeInput}
\begin{CodeInput}
$evalues
[1]  1.717934e+02  2.104448e+01  8.189522e+00  4.193012e+00  1.933547e+00
[6]  1.933547e+00  1.447649e+00  ...
$evectors
        [,1]         [,2]          [,3]          [,4]       ...     ...     
[1,]  0.008226254 -0.034169390 -0.146711415  0.0645862074   ...     ...
[2,]  -0.168561065 -0.107815274 -0.174786643 -0.2614510539  ...     ... 
       ...            ...            ...           ...      ...     ...
 [ reached getOption("max.print") -- omitted 51 rows ]
\end{CodeInput}
\end{CodeChunk}

Beside the main function, \code{npsdr\_x()} computes the estimated sufficient predictors $\hat{\phi}(\bx)=\hat{\bV}_n^{\top}\left\{ \psi_1(\bx), \ldots,  \psi_b(\bx)\right\}^{\top}$ in \eqref{eqn:nonlinear_model} for a given $\bx$.  
The usage of function \code{npsdr\_x()} is given below.
\begin{CodeChunk}
\begin{CodeInput}
 R> set.seed(200)
 R> new.x <- matrix(rnorm(n*p, 0, 1), n, p)
 R> new.y <- 0.5*sqrt((new.x[,1]^2+new.x[,2]^2))*(log(new.x[,1]^2+new.x[,2]^2))
 +          + 0.2*rnorm(n)
 R> reduced_data <- npsdr_x(object = obj_kernel, newdata = new.x, d = 2) 
\end{CodeInput}
\end{CodeChunk}
The arguments \code{object} is the object from the result of \code{npsdr()}, and \code{newdata} is a new data matrix. The argument \code{d} is for the number of sufficient predictors and its default value is 2.
In the final line of the code above, \code{npsdr\_x()} computes $\hat\phi(\bx)$ the nonlinear dimension reduction of \code{new.x} under \eqref{eqn:nonlinear_model} using kernel PSVM estimates.

\begin{figure}[!ht]
     \begin{subfigure}[b]{0.48\textwidth}
         \centering
        \includegraphics[width=1\textwidth]{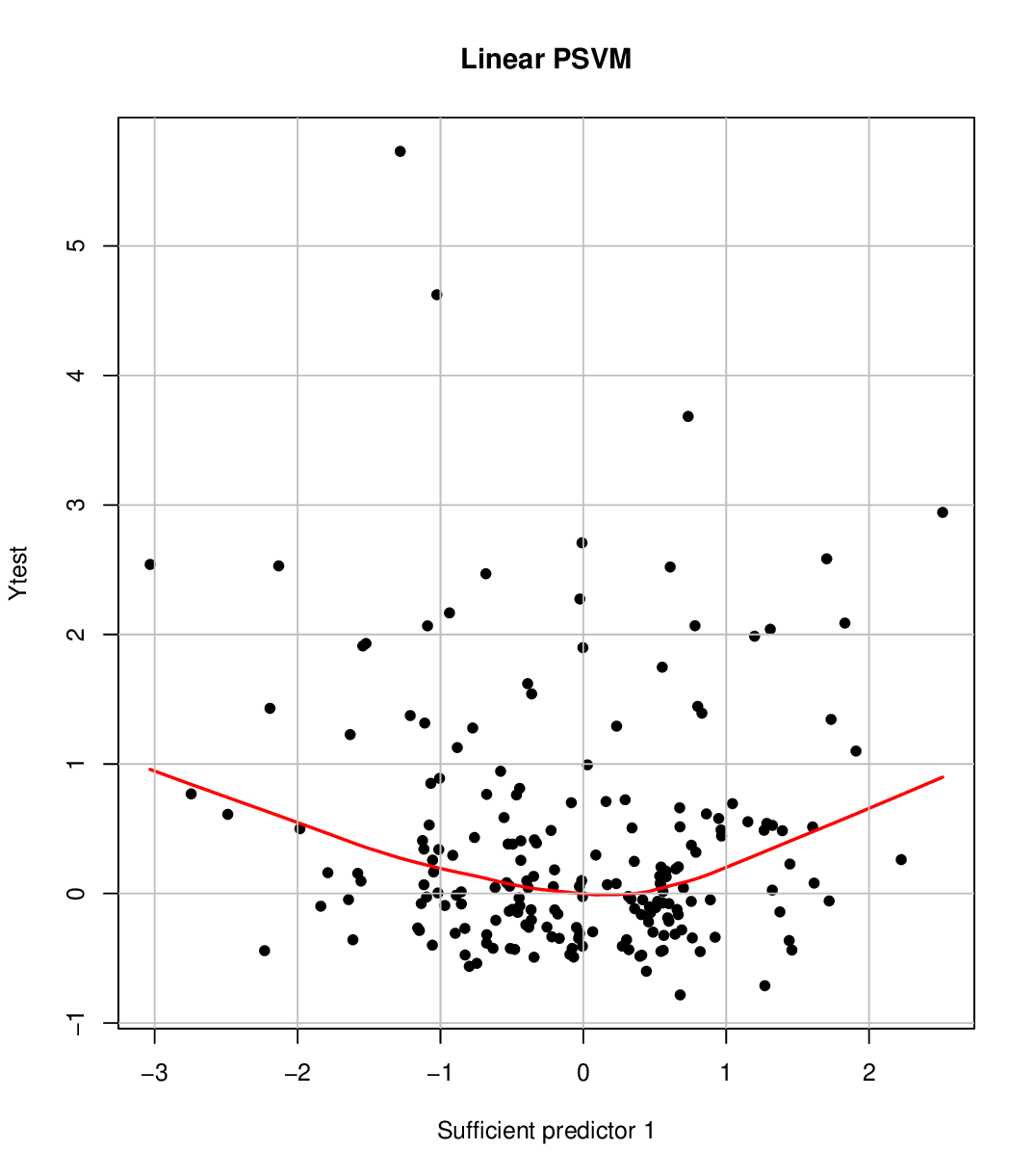}
        \caption{Linear PSVM}
     \end{subfigure}
     \hfill
     \begin{subfigure}[b]{0.48\textwidth}
         \centering
        \includegraphics[width=1\textwidth]{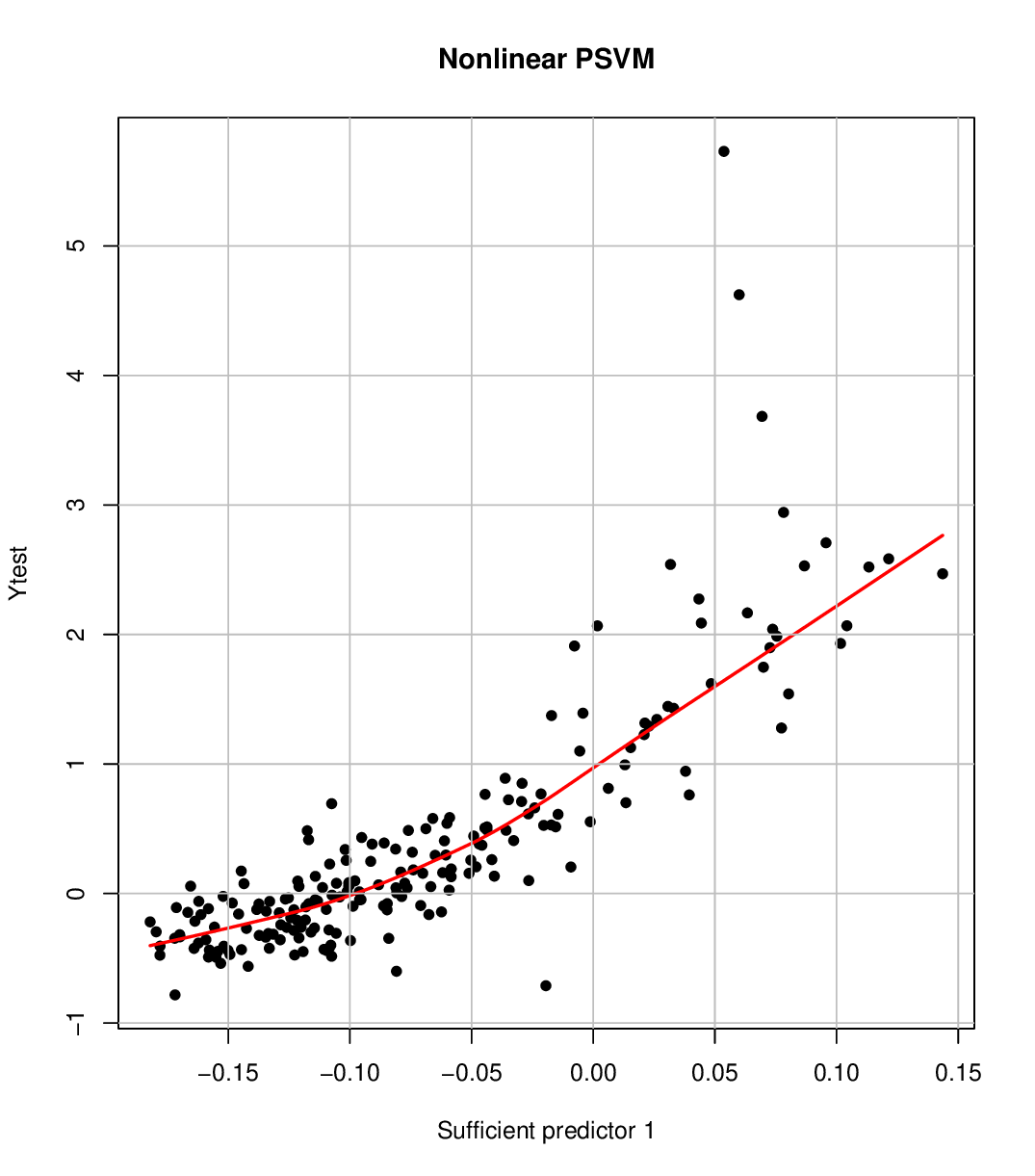}
        \caption{Kernel PSVM}
     \end{subfigure}
        \caption{Scatter plots of the response and the first sufficient predictor estimated by (a) the linear PSVM and (b) the kernel PSVM.} 
        \label{fig:nonlinear}
\end{figure}

Note that the regression function \eqref{model_regression_2} is symmetric about the origin.  
In such a scenario, linear PMs do not work and the kernel PM would be an attractive alternative. Figure~\ref{fig:nonlinear} depicts the scatter plots of the response and the first sufficient predictor estimated by (a) the linear PSVM and (b) the kernel PSVM. As expected, the linear PSVM fails to find the central subspace when the regression function is symmetric about the origin, while the nonlinear PSVM still efficiently identifies sufficient predictors.

%\begin{figure}[!htb]
%    \centering
%    \subfigure[]{\includegraphics[width=0.40\textwidth]{fig_psvm_new.eps}}
%    \subfigure[]{\includegraphics[width=0.40\textwidth]{fig_kpsvm_new.eps}}
%        \caption{Scatter plots of the response and the first sufficient predictor estimated by \textbf{(a)} the \textbf{linear PSVM} and \textbf{(b)} the \textbf{kernel PSVM}.} 
%        \label{fig:nonlinear}
%\end{figure}
%

\subsection[Functions for `rtpsdr']{Functions for \code{`rtpsdr'}}\label{sec:class_rtpsdr}
The package \pkg{psvmSDR} also provides \code{rtpsdr()} for the realtime SDR using squared loss \citep{artemiou2021real, jang2023principal}, and its usage is given below.
\begin{Code}
R> rtpsdr(x, y, obj = NULL, h = 10, lambda = 1)
\end{Code}
\begin{itemize}[itemsep=.3mm]
    \item \code{x}: covariate matrix of a new data.
    \item \code{y}: response vector of a new data.
    \item \code{obj}: the output object from \code{rtpsdr()} for the old data that contains the PM solutions and matrices for the old data. If it is not given, \code{psdr()} with the squared loss (PLSSVM) is to be applied to the given data, \code{y} and \code{x} to obtain the PM solutions and the matrices. The PWLSSVM is applied if \code{y} is binary.
    \item \code{h} a number of slices. The default is set to 10.
    \item \code{lambda} hyperparameter for the loss function. The default is set to 0.1.
\end{itemize}

\code{rtpsdr()} returns an object S3 class \code{‘psdr’} that includes
\begin{itemize}
%\item \code{M}: Working matrix
\item \code{evalues}: eigenvalues of the estimated working matrix.
\item \code{evectors}: eigenvectors of the estimated working matrix.
\item \code{r}: the PLSSVM/PWLSSVM solutions ${\mathbf{r}}$ in \eqref{eqn:rtpsdr_coef}. 
\item \code{A}: updated matrix $\bA$ for the realtime update.  
\end{itemize}

A returned object from \code{rtpsdr()} stores the state of the algorithm at each iteration $t$. 
This object is passed to the function as an argument and is returned at each iteration $t+1$ containing the state of the model parameters at that step.

The following command is an example of \code{rtpsdr()} under the streamed data scenario. The data are collected in a batch-wise manner with the batch size of $m = 500$. 
\begin{CodeChunk}
\begin{CodeInput}
 R> set.seed(1234)
 R> p <- 5
 R> m <- 500 # batch size
 R> N <- 10  # number of batches
 R> obj <- NULL
 R> for (iter in 1:N){
  + 	x <- matrix(rnorm(m*p), m, p)
  + 	y <-  x[,1]/(0.5 + (x[,2] + 1)^2) + 0.2 * rnorm(m)
  + 	obj <- rtpsdr(x = x, y = y, obj = obj) # Real time Update
  + }
\end{CodeInput}
\begin{CodeInput}
 R> round(obj$evectors, 3)
       [,1]   [,2]   [,3]   [,4]   [,5]
[1,]  1.000  0.011  0.014 -0.012 -0.002
[2,]  0.011 -0.999 -0.001 -0.033  0.004
[3,] -0.003 -0.009  0.674  0.336  0.658
[4,] -0.017  0.014  0.715 -0.521 -0.466
[5,] -0.006  0.029 -0.186 -0.784  0.592
 \end{CodeInput}
\end{CodeChunk}

For binary classification, \code{rtpsdr()} computes PWLSSVM solution \citep{jang2023principal}.
We note that \code{rtpsdr()} returns \code{psdr} object and thus \code{print.psdr()} and \code{plot.psdr()} are applicable.

\section[Summary]{Summary and discussion}
\label{sec:CON}
In this article, we introduced and tutored the \proglang{R} software package \pkg{psvmSDR}, by unifying linear and nonlinear SDR under the principal machine (PM) framework.
The package emphasizes its practical and convenient utilities through simple and user-friendly code syntax and access to various PMs.
The use of gradient descent for computation ensures efficiency and scalability, particularly beneficial in large-scale datasets.
Moreover, the package offers realtime update scheme for batch-wise data collection scenarios.
The \pkg{psvmSDR} invites researchers and practitioners in \proglang{R} communities to explore and use the powerful tools for the principal sufficient dimension reduction in their work.

\section*{Computational details}
The results in this paper were obtained using \proglang{R} 4.2.1 with the \pkg{psvmSDR} package. It is possible to have different simulation results depending on which linear algebra package is installed and which version of \proglang{R} is used.

\clearpage

\bibliographystyle{dcu}
\bibliography{references.bib}
\end{document}

%% file: tbl_loss.tex
\begin{table}[!tb]\onehalfspacing
    \centering
    \resizebox{\textwidth}{!}{
    \begin{tabular}{clcllc} \hline
                    Type & Method   & Response & Margin (Residual)   & Loss & \code{loss} \\ \hline
         \multirow{4}*{ RPM } &   SVM & \multirow{4}*{
         $\begin{array}{cc}
         Y \in \mathbb{R}\\
         \tilde{Y}_k = \mathds{1}\{Y \ge c_k\} \\
         \qquad~~ - \mathds{1}\{Y < c_k\}
         \end{array}$}     &   \multirow{4}*{$m_k = \tilde Y_k f$}  & ${[1-m_k]}_+$ & \code{`svm'} \\ 
                         & Logistic  &      &               & $\log(1 + e^{-m_k})^{-1}$ & \code{`logit'}\\ 
                        & $L_2$-SVM    &    &                & $\{[1-m_k]_+\}^2$ & \code{`l2svm'}\\ 
                         & LS-SVM &      &                & $[1-m_k]^{2}$ & \code{`lssvm'}\\ \hline 
                         %& LUM       &      &                 &$V(m_k)$  & \code{`lum'}\\ 
         \multirow{6}*{LPM}& WSVM    & \multirow{4}*{${Y = \tilde Y_k \in \{-1,1\}}$}    &  \multirow{4}*{$m = Yf$}        & $\pi_{k}(Y)[1-m]_+$ & \code{`wsvm'}\\ 
                         & Wlogistic  &      &               & $\pi_{k}(Y)\log(1 + e^{-m})^{-1}$ & \code{`wlogit'}\\ 
                        & W$L_2$-SVM    &    &               & $\pi_{k}(Y)\{[1-m]_+\}^2$ & \code{`wl2svm'} \\ 
                         & WLS-SVM &      &                & $\pi_{k}(Y)[1-m]^{2}$ & \code{`wlssvm'}\\ \cline{2-6} 
                         %& WLUM       &      &                 &$\pi_{k}(Y)V(m)$  & \code{`wlum'}\\ 
                         & Quantile  & \multirow{2}*{${Y = \tilde Y_k \in \mathbb{R}}$}    &\multirow{2}*{$r = Y - f$}  & $r \{c_k - I(r<0)\}$& \code{`qr'}\\
                         &Asym. LS  &  &    & $r^2 \{c_k - I(r<0)\}$ & \code{`asls'}\\ \hline
    \end{tabular}} \\
    \caption{Different types of the convex loss functions available in the package \pkg{psvmSDR}. The column \code{loss} indicates a specific syntax of the argument that is passed to \code{psdr()} and \code{npsdr()} to implement the corresponding methods.}
    \label{tbl:loss}
\end{table}

%% file: tbl_overview.tex
\begin{table}[!htbp]
\begin{tabular}{m{1.8cm}m{8.5cm}m{1.8cm}m{2.8cm}N}
\hline
Function & Description                                                                & Class                   & \begin{tabular}[c]{@{}c@{}}Compatible\\  methods\end{tabular}                        \\ \hline
\code{psdr()}   & \begin{tabular}[l]{@{}l@{}}Basic function for applying linear PMs\\ with a various loss functions\end{tabular}       & \multirow{4}{*}{\code{`psdr'}} & \multirow{3}{*}{\begin{tabular}[c]{@{}c@{}} \code{psdr\_bic()} \\ \code{plot.psdr()}\\ \code{print.psdr() } \end{tabular}} \\ \cline{1-2}
\code{rtpsdr()} & \begin{tabular}[l]{@{}l@{}}Real time sufficient dimension reduction through\\principal (weighted) least squares SVM\end{tabular} &                         &                                                                                     \\ \hline
\code{npsdr()}  & \begin{tabular}[l]{@{}l@{}}Basic function for applying kernel PMs \\ with a various loss functions\end{tabular}                                         &\multirow{2}{*}{\code{`npsdr'}}                    & \begin{tabular}[c]{@{}c@{}} \code{npsdr\_x()}\\ \code{plot.npsdr()}\\ \code{print.npsdr()} \end{tabular}                  \\ \hline
\end{tabular}
\caption{Main functions and compatible methods in the \pkg{psvmSDR} package.}\label{tbl:ft_list}
\end{table}